\documentclass[useAMS,usenatbib]{mnras}
   
\usepackage{graphicx}
\usepackage{aasmacros}
\usepackage{color}
\usepackage{amsmath}
\usepackage{enumitem}

\def\hmpc{$h^{-1}$Mpc}
\def\hkpc{$h^{-1}$kpc}

\def\msol{M$_\odot$}
\def\hmsol{$h^{-1}$M$_\odot$}

\def\s8{\sigma_8}

\def\lcdm{$\Lambda$CDM}

\def\x2{$\chi^2$}
\def\hmsol{$h^{-1}\,$M$_\odot$}

\def\NNm1{\langle N(N-1) \rangle}

\def\m_star{M_\ast}

\def\lcdm{$\Lambda$CDM}

\def\s8{\sigma_8}

\def\hmpc{$h^{-1}\,$Mpc}
\def\hkpc{h^{-1}{\rm kpc}}

\def\x2{$\chi^2$}
\def\hmsol{$h^{-1}\,$M$_\odot$}

\def\NNm1{\langle N(N-1) \rangle}

\def\p0{P_0(r)}

\def\lumoiii{L_{\rm [OIII]}}

\def\mgal{{M_{\ast}}}

\def\hmsol{h^{-1}{\rm M}_\odot}
\def\dn{{\rm D}_n4000}

\def\mhalo{M_{h}}

\def\msol{{\rm M}_\odot}

\def\lsat{L_{\rm sat}}
\def\lexp{L_{\rm exp}}

\def\hkpc{h^{-1}{\rm kpc}}

\def\hmpc{h^{-1}{\rm Mpc}}
\def\lcdm{\Lambda{\rm CDM}}
\def\chibar{\bar{\chi}}

\title[Halos of AGN]{On the Dark Matter Halos of Optical and
  IR-selected AGN in the Local Universe}
\author[Alpaslan \& Tinker]{\parbox{\textwidth}{ Mehmet Alpaslan, Jeremy
    L. Tinker}\\
  \footnotesize
  \\
 Center for Cosmology and Particle Physics, Department of
  Physics, New York University, New York, NY, 10003, USA\\}

\hypersetup{draft}
\begin{document}
\label{firstpage}
\pagerange{\pageref{firstpage}--\pageref{lastpage}}
\maketitle

\begin{abstract}

  We use the technique of total satellite luminosity, $\lsat$, to
  probe the dark matter halos around active galactic nuclei in the
  SDSS Main Galaxy Sample. Our results focus on galaxies and AGN that
  are the central galaxy of their halo. Our two AGN samples are
  constructed from optical emission-line diagnostics and from WISE
  infrared colors. Both optically-selected and WISE-selected AGN have
  $\lsat$ values twice as high as non-active galaxy samples when
  controlling for stellar mass and mean stellar age. This implies that
  the halos are twice as massive, but we cannot rule out that the
  increase in $\lsat$ is due to these AGN residing in younger halos at
  the same mass. When only controlling for host galaxy stellar mass,
  WISE-selected AGN also have higher $\lsat$ values than optical AGN
  at the factor of two level, consistent with previous results
  comparing the clustering of obscured and unobscured AGN. However,
  controlling for stellar age in the two populations of host galaxies
  removes half of this difference, attenuating the statistical
  significance of the difference. We perform permutation tests to
  quantify the different in the halo populations of each sample. The
  difference in star formation properties does not fully explain the
  difference in the two AGN populations, however. Although AGN
  luminosity correlates with mean stellar age, the difference in
  stellar age between the WISE and optical samples cannot be fully
  explained by differences in their AGN luminosity distributions.

\end{abstract}


\section{Introduction}

Since the discovery of active galactic nuclei (AGN), the myriad
classifications of AGN have expanded as data from disparate
wavelengths have become available. But as the taxonomy of
observational categories has expanded, progress has been made in
unifying these varying observable properties into physical models of
AGN activity (e.g., \citealt{antonucci:93, netzer:15}). The unified
model of AGN posits the existence of a dusty torus around the
supermassive black hole, the presence of which makes the observed
properties of an AGN highly dependent on the inclination angle of the
observation, even though the physical properties are the same. This
model provides a natural explanation for two populations of AGN, one
that exhibits minimal dust attenuation and is thus detected through
optical data, and one of highly obscured systems that are identifiable
at infrared (IR) wavelengths.

The advent of large IR surveys, most notably the {\it Wide-Field
  Infrared Survey Explorer} (WISE; \citealt{wise}), has created
statistical samples of dust-obscured AGN that presents a complementary
test of the unified model. If obscured and unobscured AGN only differ
in their inclination with respect to the observer, then these two
samples of objects should exhibit identical spatial clustering,
implying they occupy the same set of dark matter halos. Although not
all clustering studies agree, most have found a statically significant
difference in the bias\footnote{We define bias as $b^2=\xi/\xi_m$,
  where $\xi$ is the autocorrelation of the sample and $\xi_m$ is the
  correlation function of matter distribution. When measured at large
  scales, $r\ga 10$ $\hmpc$, $b$ can be converted to halo mass
  assuming linear bias and a model of how linear bias scales with
  $\mhalo$ (e.g., \citealt{sheth_tormen:99, tinker_etal:10_bias,
    aemulus4}). Also when measured at large scales, this $b$ is
  equivalent to that measured using cross-correlations of the tracer
  sample with the matter instead of auto-correlations.} $b$, of
dust-obscured and unobscured AGN samples such that obscured AGN have
higher clustering and thus occupy higher masses dark matter halos
(\citealt{hickox_etal:11, donoso_etal:14, dipompeo_etal:14,
  dipompeo_etal:16, mendez_etal:16,
  dipompeo_etal:17}). \cite{dipompeo_etal:17} used WISE data and
Planck CMB data to measure both quasar clustering and CMB lensing by
quasars. They found that obscured AGN have higher clustering than
unobscured at the $4\sigma$ level, yielding a factor of three
difference in the inferred halo masses.

\cite{mendez_etal:16} used PRIMUS and DEEP2 data at $z\sim 0.7$, and
thus spectroscopic redshifts, to analyze a smaller sample of AGN at
lower luminosities. \cite{mendez_etal:16} found no statistical
difference in the clustering of obscured and unobscured AGN, albeit
with errors larger than the stated difference in halo masses found by
\cite{dipompeo_etal:17}. Due to the lower AGN luminosity,
\cite{mendez_etal:16} was able to control for sample selection effects
that plague comparisons between different AGN classes. More generally,
their bias differences for samples of X-Ray, IR, and radio-selected
AGN were consistent with the differences in the host galaxy stellar
masses and star-formation properties.

In the local universe, clustering studies of X-ray selected AGN have
produced conflicting results. \cite{cappelluti_etal:10} measured AGN
autocorrelations, finding a lower clustering amplitude for obscured
sources relative to unobscured. However, \cite{krumpe_etal:18},
cross-correlating AGN sources with 2MASS galaxies, find no evidence
for a difference in large-scale clustering amplitude. They did find
that the overall AGN sample was less clustered than the reference
galaxy sample, but this sample did not control for selection
effects. \cite{li_etal:06} cross-correlated AGN identified from SDSS
spectra with reference SDSS galaxy catalogs. These samples did control
for AGN selection effects, and still found that AGN were less
clustered than the general galaxy population.

In this paper, we present a probe of dark matter halos around AGN that
is complementary to the clustering tests. Using a sample of central
galaxies\footnote{We define a central galaxy as the galaxy located at
  the center of the host dark matter halo. Satellite galaxies are
  those located within the host halo but not at the center. A host
  halo is one that is not located within the virial radius of any
  other halo.}  in the Main Galaxy Survey (MGS) of the Sloan Digital
Sky Survey (\citealt{strauss_etal:02}), \cite{tinker_etal:19_lsat} and
\cite{alpaslan_tinker:19} cross-correlated the spectroscopic MGS
central galaxies with faint imaging data to measure the total amount
of luminosity in satellite galaxies around them. Hierarchical
clustering in $\lcdm$, combined with simple prescriptions for mapping
galaxies onto dark matter halos, predicts that this quantity, $\lsat$,
should scale strongly with host halo mass
(\citealt{tinker_etal:19_lsat}). Although it is not a direct observable
of the gravitational potential around a galaxy, like weak lensing or
satellite kinematics, $\lsat$ is an indirect probe of the halo that
affords a much higher signal-to-noise measurement than direct
observables. Thus this quantity is more effective with small samples
of galaxies, such as AGN.

$\lsat$ is complementary to clustering in multiple ways. First, it is
simply an independent quantity uncorrelated with the large-scale two-point
correlation function. By the nature of our sample construction, we
only analyze central galaxies, removing satellite galaxies from
consideration. This makes the interpretation of the inferred halo
masses more clear. Second, $\lsat$ breaks degeneracies with halo
properties that affect clustering apart from halo mass. The implicit
assumption in the clustering studies cited above is that the
connection between AGN properties and dark matter halos is confined
only to the {\it mass} of the halo---i.e., more massive halos cluster
more strongly, thus differences in $b$ imply differences in
$\mhalo$. However, it is well-known that halo clustering is highly
dependent on other properties apart from mass. This effect is referred
to as `halo assembly bias (see, e.g., the review
by \citealt{wechsler_tinker:18} and references therein). For example,
at fixed $\mhalo$, early-forming halos show significantly enhanced
clustering relative to their late-forming counterparts. Thus, the
enhanced clustering of dust-obscured AGN may be driven by the
formation history of the halos they occupy, rather than the halo
mass. This would still support the interpretation that dust-obscured
AGN are distinct from optical AGN, but the connection with the dark
sector---and thus possible evolutionary scenarios---is markedly
different.

The connection to $\lsat$ is that, in addition to impacting
clustering, early-forming halos have significantly less substructure
as well. For example, if the elevated bias in \cite{dipompeo_etal:17}
is to due assembly bias, $\lsat$ for dust-obscured AGN would be
suppressed relative to optical samples, and suppressed relative to the
overall sample of non-AGN galaxies. On the other hand, if the enhanced
clustering is truly due to the halo mass, the $\lsat$ would be
enhanced as well, by roughly the same factor as the $\mhalo$.

We will show $\lsat$ results for three different samples of central
galaxies from the MGS: the overall sample, which we will use as a
control sample, a sample of optical AGN within the MGS selected
using spectral line diagnostics, and a sample of dusty AGN
selected from the MGS using broadband colors from WISE data. Like
\cite{mendez_etal:16}, we will control for sample selection effects
the estimated halo masses. In \S 2 we present our data, including the
sample of central galaxies, with relevant background on
how we measure $\lsat$. In \S 3 we present the $\lsat$ results for the
different samples of galaxies. Due to the limited size of these
samples, we focus on the statistical significance of the result,
presenting multiple ways to analyze the results. In \S 4 we will
compare our results to previous data and discuss them in the context
of the unified AGN model. For all distance calculations we assume a
flat, $\lcdm$ cosmology with $\Omega_m=0.3$.

\section{Data}
\label{s.data}

This analysis is based on the sample of MGS central galaxies and
$\lsat$ measurements from \cite{alpaslan_tinker:19}. This sample is
constructed from the NYU Value-Added Galaxy Catalog (NYU-VAGC;
\citealt{blanton_etal:05_vagc}). Specifically, we use the magnitudes,
redshifts, and positions from the {\tt dr72bright34} sample. Derived
galaxy quantities, specifically $D_n4000$ and AGN classification, are
obtained from the MPA-JHU value-added catalog\footnote{\tt
  https://www.sdss3.org/dr10/spectro/galaxy\_mpajhu.php}. The MPA-JHU
catalog corresponds to the DR8 SDSS public data release, which
contains all galaxies in the DR7 sample. Additional photometric data
is from the DECALS Legact Imaging Surveys (DLIS;
\citealt{legacy_surveys}), which is a combination of DECam data at
declination~$\le 30^\circ$ and Kitt Peak data at higher
declination. Although the current DLIS data release is DR8 (NB, this
numbering scheme is distinct from the SDSS DR numbers), the
\cite{alpaslan_tinker:19} results use DR6 (from Kitt Peak) and DR7
(from DECam) results. The DR6+DR7 combined photometry covers roughly
75\% of the final MGS footprint. We detail our samples, and how these
data are used, further in this section.

\subsection{Total Satellite Luminosity}

Total satellite luminosity is a method for probing the relationship
between galaxies and halos. The amount of substructure within a dark
matter halo is roughly self-similar with the mass of the halo. These
substructures will contain faint satellite galaxies, thus the total
amount of light in satellite galaxies is a strong probe of the dark
matter halo around the galaxy that sits at the center of that
halo. Most satellite galaxies are below the magnitude limit of the MGS
spectroscopic sample, thus the $\lsat$ measurements require using
deeper imaging data. In \cite{tinker_etal:19}, we used DLIS data as
our source of imaging due to its increased depth and better resolution
over legacy SDSS imaging. The typical $r$-band depth of the DLIS data
is $r\approx 24$, as compared to the MGS spectroscopic limit of
$r=17.7$. To measure $\lsat$, we sum the total $r$-band light from
imaging galaxies, excluding the spectroscopic central galaxy, within a
50 $\hkpc$ (comoving) aperture. To estimate what fraction of those
galaxies are associated with the halo, as opposed to random
interlopers projected along the line of sight, we measure the expected
interloper contribution from randomly placed apertures of the same
size. For both central galaxies and randomly placed apertures, we
require that there are no other MGS spectroscopic galaxies within the
aperture. After subtracting off the background contribution for a
given central galaxy, we calculate luminosity by assuming that all
objects are at the redshift of the central galaxy. We limit the
$\lsat$ value to satellites brighter than $M_r-5\log h=-14$. We refer
the interested reader to \cite{tinker_etal:19_lsat} for a
comprehensive description of the measurements and tests of the method.

Measurements of $\lsat$ for a single galaxy are, of course, too noisy
to be of much use. A significant fraction of galaxies have negative
values of $\lsat$ simply due to random variation in the background
galaxy counts, as well as fluctuations in the intrinsic number of
satellites in a give halo. Thus, we stack central galaxies and take
the average $\lsat$ values. In \cite{alpaslan_tinker:19}, we
demonstrated that $\lsat$ correlates strongly with $\mgal$, as
expected from the correlation between $\mgal$ and
$\mhalo$. At fixed $\mgal$, $\lsat$ also correlated with other galaxy
properties, implying that these properties carry information about the
halos they live in, beyond the information encoded in their stellar
masses. \cite{alpaslan_tinker:19} used additional data---the
large-scale density around the satellite galaxies--- to break this
degeneracy between $\mhalo$ and halo assembly bias. Here, due to lack
of statistics to probe environments, we focus on the $\lsat$
measurements themselves as a test of whether the halo around different
types of AGN are distinct from each other and distinct from the
overall non-AGN population.

\subsection{Central Galaxies}

To characterize MGS galaxies we use stellar masses using the PCA
method of \cite{chen_etal:12}. This is the same stellar mass
definition employed in \cite{tinker_etal:19_lsat} and
\cite{alpaslan_tinker:19}.  All analyses in this paper are performed on
galaxies that are inferred to be central within their dark matter
halo. Although there is a population of AGN within the set of
satellite galaxies (\citealt{pasquali_etal:09}), $\lsat$ measurements
are much more difficult to interpret for satellites, as the number of
faint galaxies around a satellite is more a function of host halo and
the satellite's position within it. For central galaxies, $\lsat$ is
direct probe of the host dark matter halo.

Methods to determine whether or not a galaxy is a central will always
yield a sample with some amount of incompleteness and
impurities. Group catalogs on volume-limited samples can construct
sets of central galaxies with both high completeness and purity
(\citealt{tinker_etal:18_p2}), but requiring that the galaxy sample be
volume-limited is highly restrictive. Given the low frequency of AGN,
being able to analyze the largest possible parent sample of galaxies
is paramount. In this paper we use the `central galaxy finder'
introduced in \cite{tinker_etal:19_lsat}, and used in
\cite{alpaslan_tinker:19}. In tests using a mock flux-limited MGS-like
sample, and using a restrictive criteria to minimize impurities, this
method yields a sample of central galaxies that is 93\% pure and 72\%
complete. The small amount of satellite galaxies that make it into
this `pure' central sample do impact the average $\lsat$ measured in
bins of central galaxy stellar mass; at $\mgal\la 10^{10.5}$ $\msol$,
impurities increase $\lsat$ by 0.1-0.2 dex, while for more massive
galaxies the impacts are nearly negligible. Because AGN have roughly
the same satellite fraction as the overall galaxy population
(\citealt{pasquali_etal:09}), this should not bias the comparison of
$\lsat$ between AGN samples and non-AGN samples.

\begin{figure}
  \hspace{-0.5cm}
  \includegraphics[width=9cm]{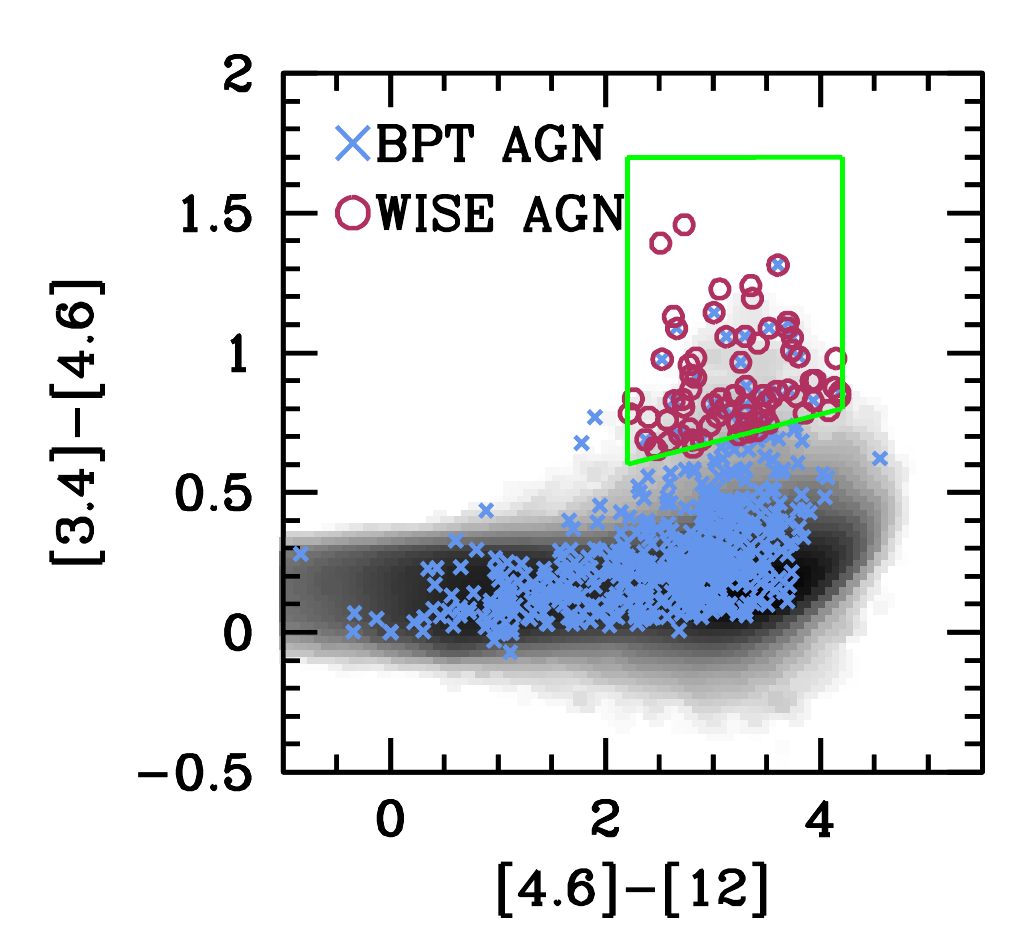}
  \vspace{-0.6cm}
  \caption{ \label{agn_selection} Selection of dust-obscured AGN in
    WISE color-color space. The axis labels indicate the WISE bands
    used in each color. The grayscale represents the number density of
    MGS galaxies at each point in color space, using a logarithmic
    color scale. The blue crosses show a random sample of galaxies
    that are identified as AGN using the BPT line diagnostics in the
    MPA-JHU catalog. The red circles are objects contained within the
    green trapezium in the upper right. These frequency of these
    objects is 3.1\% of the overall sample. These color cuts are taken
    from \citealt{jarrett_etal:11}, optimized to find dust-obscured
    AGN. The frequency of these objects is 0.8\% of the full sample.
  }
\end{figure}

\begin{figure*}
  \includegraphics[width=15cm]{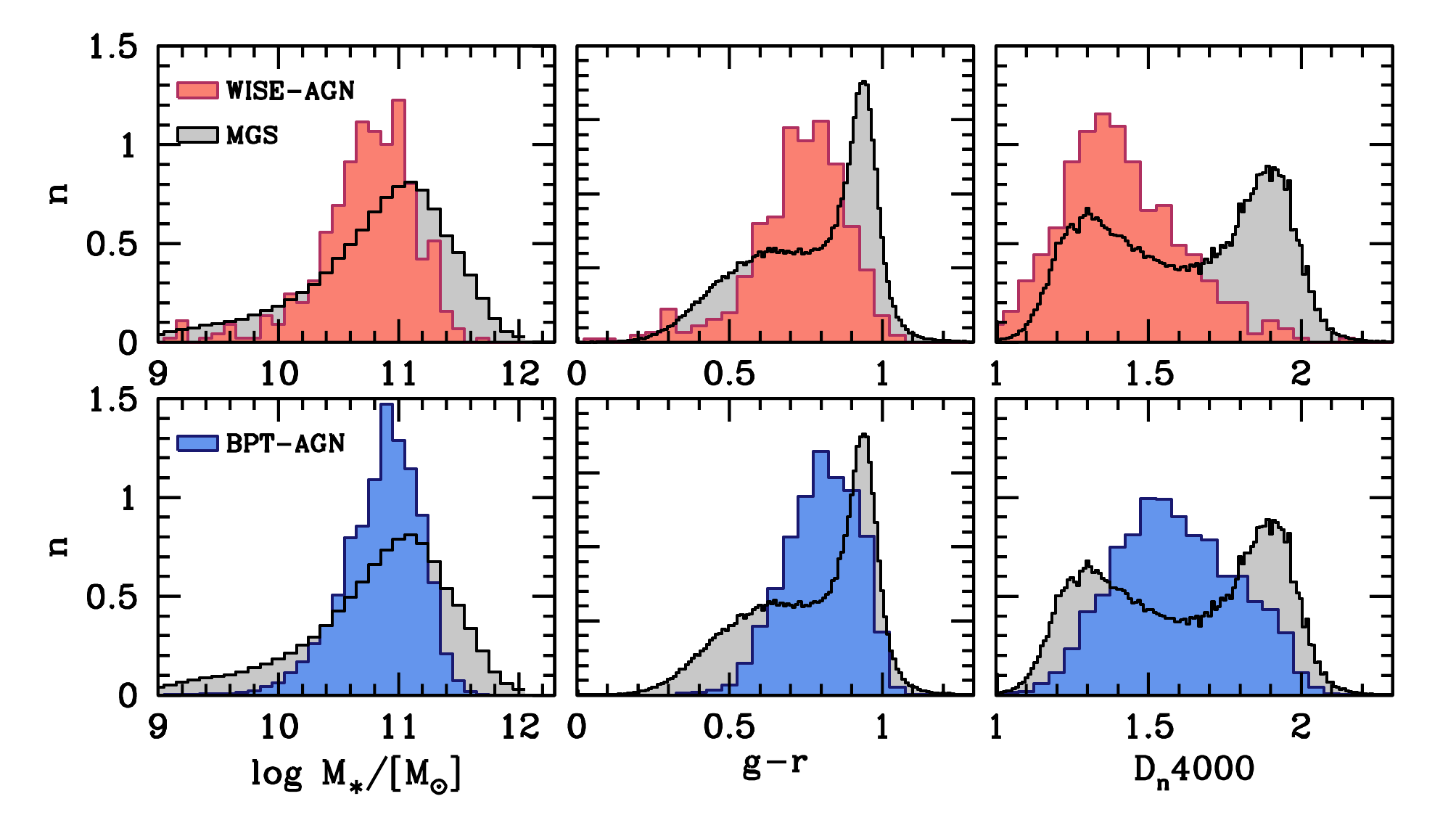}
  \vspace{-0.1cm}
  \caption{ \label{sample_properties} Properties of the AGN samples
    relative to the overall MGS sample of central galaxies. {\it Top
      Row:} Comparison of the normalized distributions, $n$, of
    stellar mass, $g-r$ color, and $D_n4000$ for the WISE-selected
    sample. {\it Bottom Row:} Same as the top row, but now for the
    distribution of BPT-selected AGN. }
\end{figure*}

\begin{figure}
  \hspace{-0.5cm}
  \includegraphics[width=9cm]{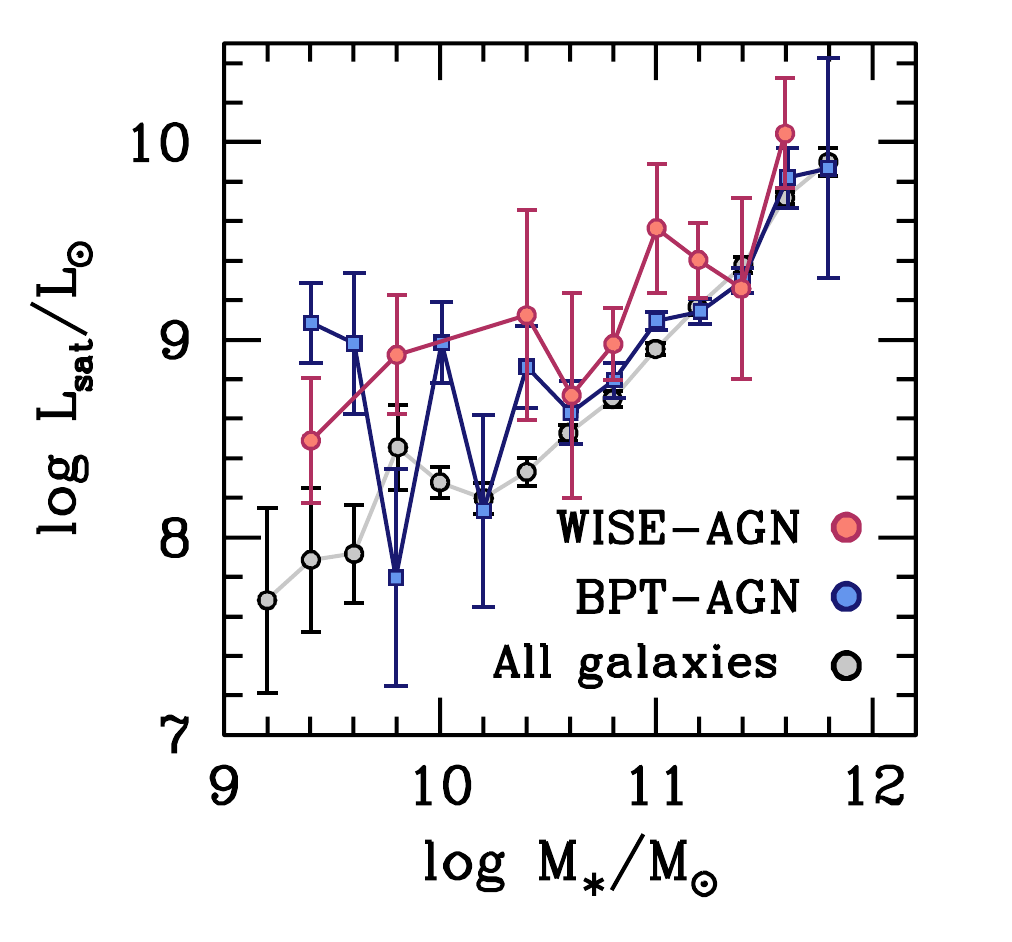}
 \vspace{-0.4cm}
  \caption{ \label{lsatbar} The mean $\lsat$ as a function of central
    galaxy stellar mass, $\mgal$. The gray connected circles show the
    relation for the full sample of central galaxies. The connected
    blue circles show the relation for the BPT-selected AGN sample. The
    connected red circles show the relation for the WISE-selected AGN
    sample. For each data point, the error bar is estimated using
    bootstrap resampling of the central galaxy samples. }
\end{figure}

\subsection{AGN Selection}

Optically-selected AGN are taken from the public MPA-JHU DR8
catalog. The selection is described in \cite{brinchmann_etal:04} and
\cite{kauffmann_etal:03}. It is based on the BPT (\citealt{bpt})
line-ratio diagnostic to separate star-forming and active
galaxies. The sample we use is a strict AGN classification, removing
galaxies with line ratios indicative of LINERS and composite
spectra. From 559,028 galaxies in the NYU-VAGC sample, 17,272 of them
match this classification. After restricting the sample to be only
`pure' central galaxies, removing objects that have MGS interlopers
along the line of sight, and choosing only galaxies that lie in the
overlap with DLIS DR6+DR7 imaging, the number of optical AGN used in
the analysis is 4,432.

Infrared-selected AGN are identified through the use of WISE
photometry available in the DLIS public data release
(\citealt{legacy_surveys}). Many studies have shown WISE color-color
analysis can effectively isolate dust-obscured AGN from the general
galaxy population (\citealt{jarrett_etal:11, yan_etal:13,
  satyapal_etal:14}) In this work, we use the unWISE co-added data,
with forced photometry for DLIS optical sources, from
\cite{lang:14}. All MGS objects are detected in the DLIS imaging where
the two surveys overlap, and matched catalogs between the DLIS data
releases and the SDSS Legacy samples are provided on the DLIS public
pages\footnote{\tt http://legacysurvey.org/}. Figure
\ref{agn_selection} shows a subsample of MGS galaxies in WISE
color-color space. Those objects that are also identified as optical
AGN using the line diagnostics are also indicated. The green trapezium
shows the AGN selection from \cite{jarrett_etal:11}. In the NYU-VAGC,
1,942 objects pass this selection. After trimming the catalog in the
same manner as for the BPT selection, the final number of WISE AGN
used in the analysis is 449.

\subsection{Host galaxy properties}

The host galaxy properties of these two AGN samples, relative to the
overall sample of MGS central galaxies, is shown in Figure
\ref{sample_properties}. To characterize these populations, we use
galaxy stellar mass, the broadband $g-r$ color, and the 4000-\AA\
break in the galaxy spectra, $D_n4000$. The $g-r$ color is sensitive
to both the age of the stellar population and the amount of dust in
the host galaxy. A value of $g-r=0.8$ is a rough break-point between
the blue cloud and red sequence. The 4000-\AA\ break is also sensitive
to the mean stellar age of the host galaxy, but is largely insensitive
dust content, making it a more efficient diagnostic for identifying
star-forming and quiescent galaxies. For the overall galaxy
population, the distribution of $\dn$ is more clearly bimodal than the
color distribution. A break-point of $\dn=1.6$, at the minimum between
the two modes, is a reasonable separation point between star-forming
and quiescent galaxy samples.

For both AGN samples the median stellar mass is $\sim 10^{10.8}$
$\msol$, with a dearth of both very massive and low-mass galaxies. The
host galaxies of BPT-AGN have redder broadband colors than the overall
sample, but the distribution is unimodal and peaks at the valley
between the two modes in the MGS distribution. Additionally, the $\dn$
values also peak at the transition point between star-forming and
quiescent populations, as shown by \cite{kauffmann_etal:03}. The
fraction of BPT host galaxies above the breakpoint in color is 52\%,
while the fraction at $\dn>1.6$ is 47\%. The $g-r$ color distribution
for WISE-AGN host galaxies is also unimodal, but the peak is shifted
slightly blue-ward of the BPT distribution. The mean sellar ages of
the WISE host galaxies, however, are clearly indicative of a sample of
actively star-forming galaxies. The fraction of WISE-AGN host galaxies
in the red sequence is 34\%, but the fraction categorized as quiescent
by their $\dn$ break is only 15\%. This is consistent with the
IR-selected AGN sample in \cite{mendez_etal:16}, which were
predominantly star-forming objects, as well as the $z\sim 0$ of
WISE-selected AGN in \cite{weston_etal:17}.

\begin{figure*}
  \includegraphics[width=16cm]{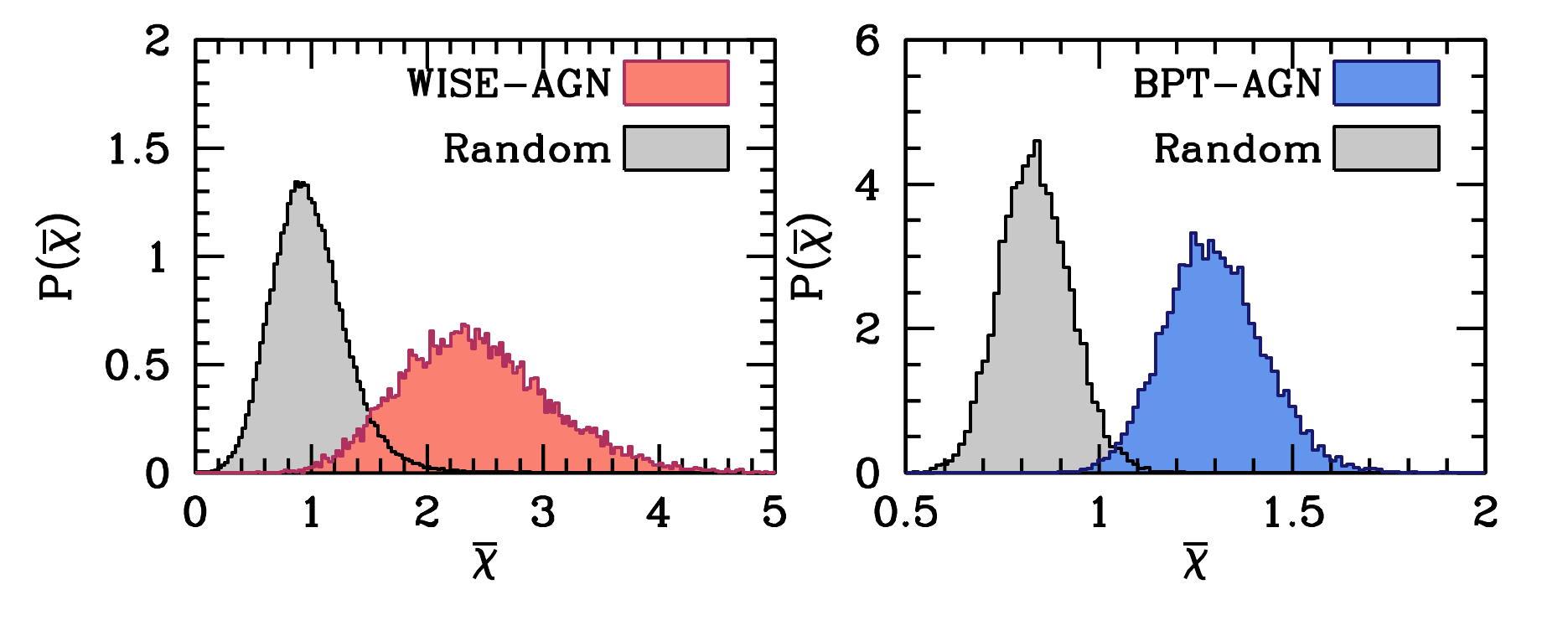}
  \caption{ \label{lxprob} The probability distribution functions of
    the quantity $\chibar$, defined as the mean of $\lsat/\lexp$. In
    both panels, the gray histograms show the distributions of
    $\chibar$ for random samples of the population of non-AGN central
    galaxies, where each random sample has the same size as the AGN
    sample. The random selection is weights such that the distribution
    of $\mgal$ in the AGN sample is matched. Note the difference in
    the $x$-axis ranges in both panels; because the BPT-AGN sample is
    $\sim 10$ times larger, the width of the $\chibar$ PDF is significantly
    smaller. In each panel, the colored histogram is the distribution
    of $\chibar$ from bootstrap resampling.}
\end{figure*}

\section{Results}

We are asking three distinct questions: Are the halos of WISE-AGN
distinct from the overall non-AGN population? Are the halos of BPT-AGN
also distinct from the overall galaxy population? And last, are the
halos of the different AGN classes distinct from each other? We will
investigate these questions in three different ways: by directly
measuring the $\lsat$-$\mgal$ relations for the three different
samples, by comparing the mean $\lsat$ of the AGN samples to the
expected mean given the stellar mass distribution of the samples, and
by use of a permutation test.

\subsection{Mean $\lsat$ }

Figure \ref{lsatbar} shows the mean $\lsat$ as a function of $\mgal$,
taken from \cite{alpaslan_tinker:19}. In each bin, the error is
calculated using 100 bootstrap resamples of the sample of central
galaxies. The mean $\lsat$ for AGN identified in the BPT sample are
shown with the blue symbols. The results are noisy at
$\mgal\la 10^{10}$ $\msol$, but within the errors the $\lsat$ values
for BPT AGN are either consistent with, or slightly higher than, the
overall sample of central galaxies. In subsequent subsections we will
make this comparison more quantitatively. The $\lsat$ values for the
WISE-AGN, however, are clearly above the mean relation. The small
sample size yields large statistical errors for the WISE-AGN
sample. Thus, if one were to calculate these two sets of measurements
using a $\chi^2$ statistic, one would find minimal difference between
the two samples. However, it is worth noting that {\it all} the values
of $\lsat$ for the WISE-AGN are above the mean relation. And if the
measurements for this sample are merely statistical fluctuations,
those fluctuations would need to go in the same direction for the vast
majority of the sample. We present our statistical tests of this
scenario presently.

\subsection{Are the $\lsat$ values different?}

Figure \ref{lxprob} shows our first quantitative test of the three
populations of central galaxies. As demonstrated in Figure
\ref{lsatbar}, $\lsat$ depends strongly on $\mgal$. However, binning
the AGN samples by $\mgal$ only reduces the numbers of objects to work
with, making it difficult to determine whether the AGN samples are
above the mean. Thus, we define a quantity
$\chi=\lsat/\lexp$, where $\lexp$ is the ``expected'' value of $\lsat$
at a given value of $\mgal$, given the mean relation in Figure
\ref{lsatbar}. For the sample of BPT-AGN, the mean value is
$\chibar = 1.29\pm 0.13$. For the sample of WISE-AGN, the mean value
if $\chibar = 2.42 \pm 0.64$. For both these measurements, the error
in the mean is estimated by bootstrap resampling on the AGN
sample. The bootstrap distributions of $\chibar$ for both BPT and WISE
samples are shown in Figure \ref{lxprob}. From the bootstrap analysis
alone, WISE-selected AGN live in halos that have $\lsat$ values
$1.88\pm 0.52$ higher.

From the host galaxy properties in Figure \ref{sample_properties}, we
note that $\mgal$ is not the only galaxy parameter that may have an
impact on $\lsat$ and the expected halo mass. Using weak lensing
\cite{mandelbaum_etal:16} found strong bimodality in the halo halo
masses of massive red and blue central galaxies, with red galaxies
occupying higher mass halos. These results agree well with the $\lsat$
results in \cite{tinker_etal:19_lsat} for red and blue subsamples,
under the assumption that $\lsat$ is a function of $\mhalo$
only. Going beyond a simple red-blue split of the galaxy population,
\cite{alpaslan_tinker:19} found that $\lsat$ correlated continuously
with $\dn$ at fixed $\mgal$, with a minimum at the green valley. Given
that the peak of the $\dn$ distribution for BPT-AGN is in the green
valley, we can ask what $\chibar$ is for a random sample of central
galaxies with both the same $\mgal$ and $\dn$ distributions.

The gray histograms in Figure \ref{lxprob} show the distribution of
$\chibar$ values for random samples of the non-AGN galaxy population,
selected such to match the distributions of $\mgal$ and $\dn$, and
matching the number of objects in each sample. For the random sample
tailored to the WISE-selected sample, $\chibar$ is unbiased relative
to the the general population. The results for the BPT-like random
sample is quite different. Given their location in the green valley,
$\chibar$ for the BPT-matched random sample is significantly lower
than expected by $\mgal$ alone, with a $\chibar=0.83\pm
0.09$. Combining this with the bootstrap results of $\chibar$ for the
BPT-AGN, these AGN have $\lsat$ values $1.54\pm 0.16$ times higher
than expected from given the demographics of their host galaxies.

Thus, when controlling for both stellar mass and star formation in the
comparison of the $\chibar$ values for the WISE and BPT samples, the
ratio of WISE/BPT is $1.57\pm 0.44$. The WISE AGN live in halos with
higher $\lsat$ values, but the statistical significance of this result
is very low, at only $1\sigma$. 

\subsection{Permutation tests}

In order to assess the statistical significance of the differences
between the $\lsat$ values of the different AGN populations, we
perform a permutation test (also referred to as a randomization
test). For two given samples of observations $A$ and $B$ with means
$\bar{x}_A$ and $\bar{x}_B$ a permutation test can measure, by
examining the difference between the two sample means, the statistical
significance of the null hypothesis $H_0$ that both $A$ and $B$ are
drawn from the same parent distribution. In practice, the test is
simple to perform: the difference between both means, $T_0$, is
computed. Following this initial calculation, all observations from
$A$ and $B$ are pooled together and resampled at random into new
subsamples according to the relative sizes of both samples ($N_A$ and
$N_B$). The difference of means is calculated again, and this
procedure is repeated for every possible permutation of $A$ and $B$,
yielding a distribution of \emph{the difference of means}. Where $T_0$
lies relative to this distribution then allows one to accept or reject
the null hypothesis. Our expectation is that the results of these
tests should generally follow those from the previous subsection, but
they are a quantitatively distinct test.

In our implementation of the permutation test, we set $N_A=N_B$ and,
when comparing MGS galaxies to AGN samples, randomly sample the larger
of the two populations such that the distribution of galaxy properties
matches the sample selection of the smaller sample. We perform the
test for a total of 1,000 random samples from the larger population,
where the results have converged. Each test yields a deviate---a
location of $T_0$, in the range of [0,1], in the distribution created
by the permutations. For example, a deviate of 0.5 is exactly at the
median of the permutations, and a value of 0.99 means the null
hypothesis can be excluded at $3\sigma$ confidence. 

\begin{figure}
  \includegraphics[width=8cm]{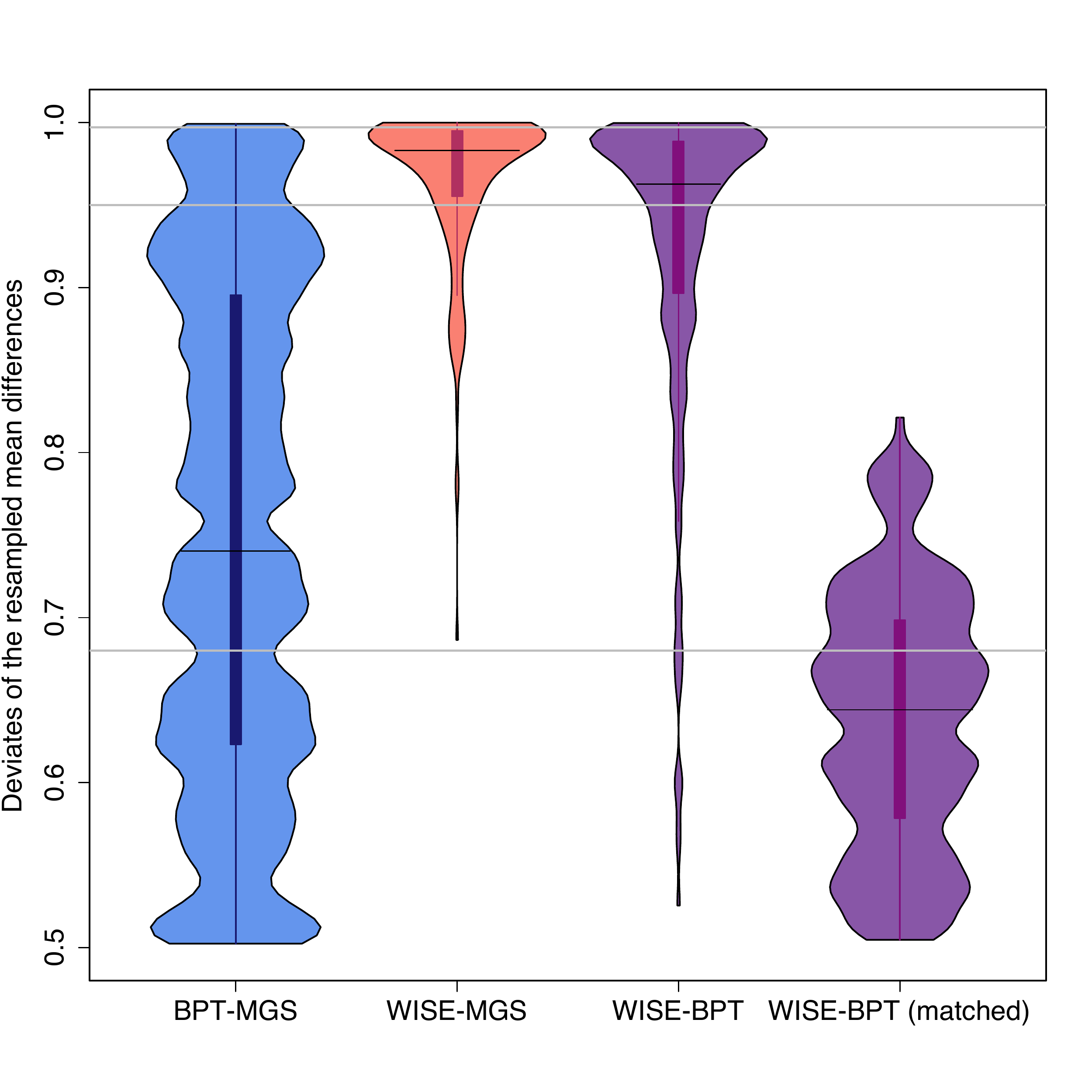}
  \vspace{-0.4cm}
  \caption{ \label{violin} Violin plots showing the distributions of
    deviates generated in 1,000 realizations of the permutation test
    for all AGN samples. See text for details on our implementation of
    the test. For the first three violins, results only control for
    stellar mass in the comparison of the distributions. The grey
    horizontal lines denote quantiles that correspond to a $1\sigma$,
    $2\sigma$, and $3\sigma$ deviate respectively. These results
    suggest that there is an almost $3\sigma$ significance in the
    permutation test result between WISE-selected and BPT-selected
    AGN, when controlling for stellar mass. The right violin (labeled
    as ``matched''), however, shows how this result is mitigated when
    controlling for $\dn$ in the comparison as well.}
\end{figure}

In Figure \ref{violin} we show `violin' plots of the results of
performing permutation tests on the different AGN samples we have
examined in this work. In the first two violins, we have sampled the
the MGS galaxies to match the distribution of stellar masses of the
AGN samples. The figure shows the distribution of deviates, with the
central thick bar displaying the inner-50\% range; the horizontal
black line displaying the median, and the full distribution being
shown as the wings of the violins. The horizontal gray lines show the
$1\sigma$ - $3\sigma$ significance values. One final consideration is
that for the purposes of our results, we can reject the null
hypothesis if the statistical significance of the difference in the
means is positive \emph{or} negative. In other words, if the full
distribution of the mean differences were normally distributed around
0 with a nonzero standard deviation, we can reject the null hypothesis
if the difference between the two means were on the positive or
negative tail of the distribution, so long as it was at or above
$3\sigma$ significance. For this reason, in Figure \ref{violin} we
only show the deviates from $0.5$ to $1$, and when performing the
actual calculation we map all deviate values below 0.5 to be between
0.5 and 1. We note that this only impacts the results for the BPT-MGS
comparison.

We will discuss the permutation results presented in Figure
\ref{violin} from left to right. For the BPT-MGS comparison, we cannot
reject the null hypothesis that these two samples have the same
distribution of $\lsat$ values, and thus the same dark
matter halos. For the WISE-MGS comparison, on the other hand, we can
reject the nll hypothesis. Greater
than 75\% of the deviates are above the 2$\sigma$ line. Additionally,
but with somewhat less confidence, we reject the null hypothesis that
the WISE and BPT samples are drawn from the same distribution. These
results are in agreement with the $\chibar$ results from Figure
\ref{lxprob}. We note again that these first three results only
control for stellar mass in the permutation tests. The last violin
shows the WISE-BPT comparison, now controlling for both $\mgal$ and
$\dn$. Here, as in the previous subsection, the significance of the
difference between the two samples is eliminated, implying that the
difference in the two sets of halos is driven mostly by the
star-forming properties of their host galaxies. 

\begin{figure}
  \includegraphics[width=8cm]{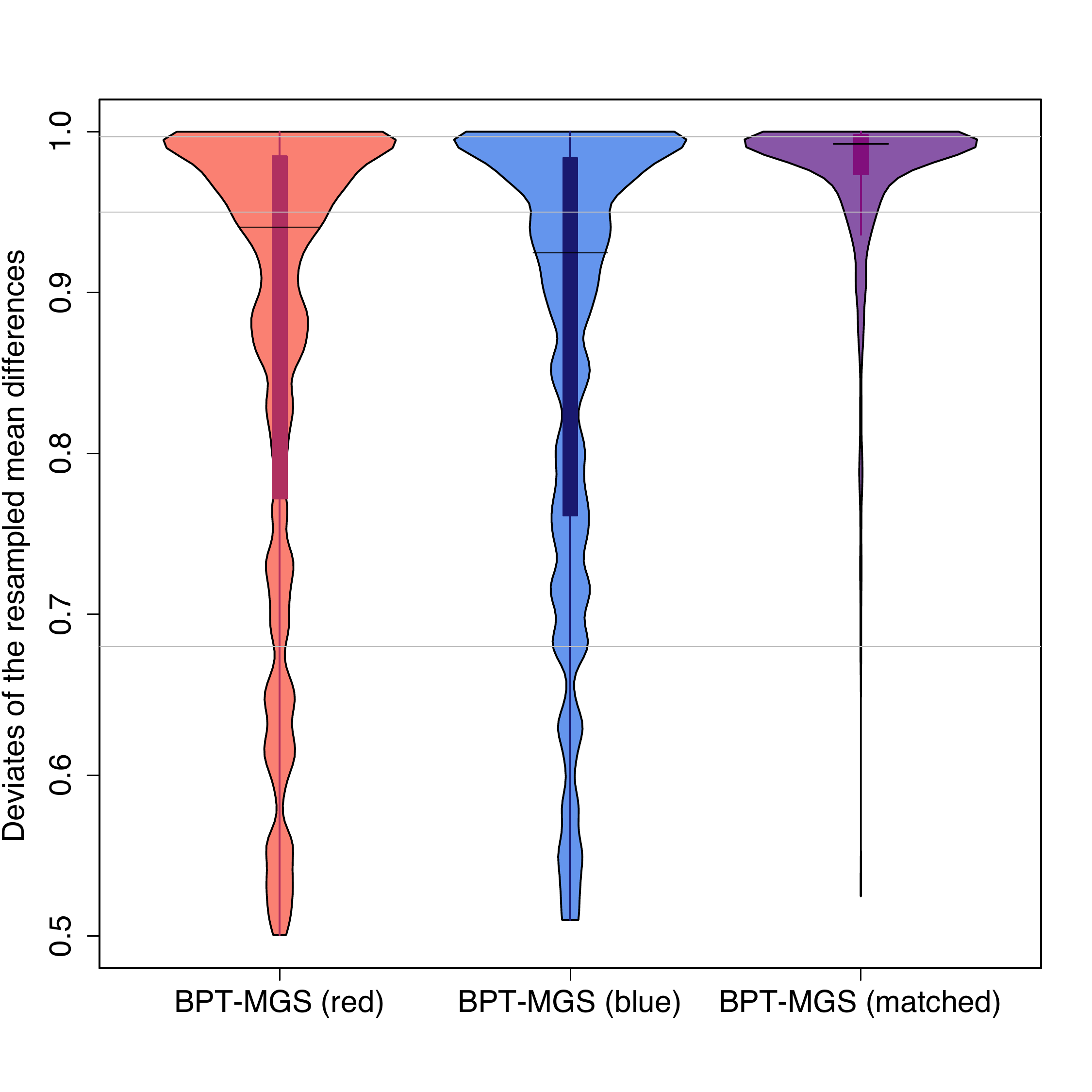}
  \vspace{-0.4cm}
  \caption{ \label{violin2} Violin plots for the comparison of the
    BPT-AGN sample to the non-AGN sample. From left to right, the
    three violins indicate the distribution of deviates for galaxies
    (and AGN host galaxies) classified as ``red'' (quiescent) by their
    $\dn$ value, galaxies classified as ``blue'' (star-forming) by
    $\dn$, and where the MGS population is sampled to match both the
    stellar masses and $\dn$ values of the BPT-AGN sample. }
\end{figure}

Figure \ref{violin2} explores the BPT-MGS comparison in more
detail. The first two violins compare the two samples, but for only
quiescent and star-forming galaxies, respectively, from both
samples. Here we use $\dn=1.6$ to separate the two samples. In
contrast to the BPT-MGS comparison in FIgure \ref{violin}, we can
reject the null hypothesis at $\sim 2\sigma$ for each subsample. The
right-most violin shows a full comparison, but now matching the
distribution of both $\mgal$ and $\dn$. In this comparison, we can
reject the null hypothesis at nearly 3$\sigma$.

\begin{figure*}
  \includegraphics[width=16cm]{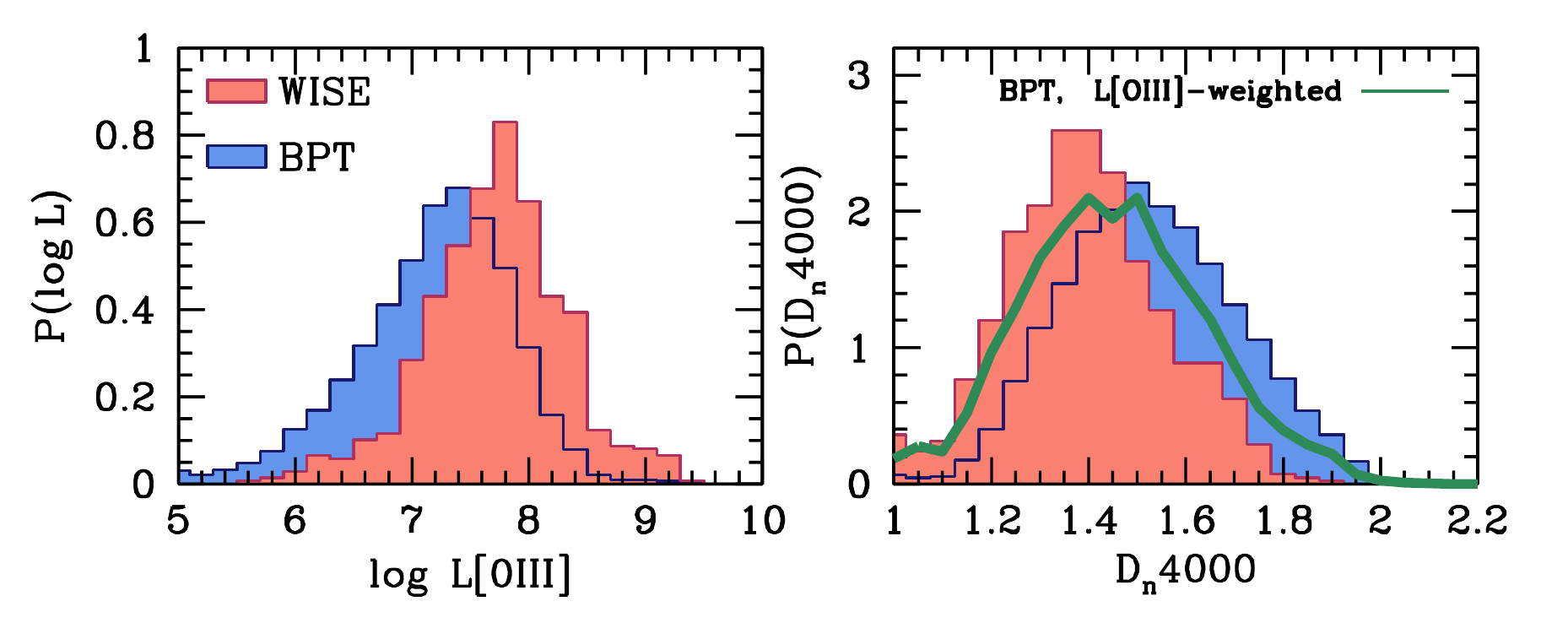}
  \vspace{-0.4cm}
  \caption{ \label{oiii} {\it left Panel:} Normalized distributions of
    $\log \lumoiii$ in MGS galaxies identified as AGN in their
    spectra. [OIII] luminosity values are given in units of solar
    bolometric luminosity, in the same manner as in
    \citealt{kauffmann_etal:03}. {\it Right Panel:} Normalized
    distributions of $\dn$ AGN. The blue histogram is all BPT AGN,
    while the red histogram are those AGN that lie within the WISE
    selection box in Figure \ref{agn_selection}. The dark green curve
    is the distribution of $\dn$ of the BPT sample, when weighted to
    match the $\lumoiii$ distribution of WISE AGN. Although the mode
    of the weighted distribution is shifted to lower $\dn$, a KS test
    yields a probability that the red histogram is drawn from the
    green distribution at $10^{-21}$.}
\end{figure*}

\section{Summary and Discussion}

In this paper we have provided multiple ways to test whether
optically-selected AGN and IR-selected AGN are part of the same
overall population of objects, and whether either either sample live
in distinct halos from the general population of galaxies.

The $\lsat$ measurements demonstrate that both samples of AGN live in
distinct halos from the general population when controlling for
stellar mass and mean stellar age. For the $\chibar$ test, the
WISE-selected AGN have higher $\lsat$ values at $\sim 2\sigma$ level,
while in the permutation test the WISE population of halos is distinct
from a random sample at nearly $3\sigma$ For the BPT-selected AGN, the
$\chibar$ test finds that the halos of these AGN have higher $\lsat$
values at $3\sigma$ significance, while the permutation test settles
in to just under $2\sigma$ confidence. We note that, when only
controlling for stellar mass and not for $\dn$, the conclusions for
the WISE-AGN sample are unchanged, but the evidence for BPT-AGN being
in different halos would be substantially weaker.

The clustering study of \cite{li_etal:06} is closest to our $\lsat$
analysis in terms of sample selection and systematic controls. They
compared the clustering of AGN in SDSS to samples of galaxies
controlled by various properties, including $\mgal$ and $\dn$. They
found that AGN were {\it less} clustered than the control samples,
more so when $\dn$ was used as part of the control sample. This is
the opposite of the signal measured here. The AGN sample they used,
however, is significantly different. They included LINERS and
composite spectra, yielding a sample significantly larger than
ours. If AGN selection is the cause of the difference, then the
additional populations in their AGN would have to be in significantly
lower mass halos than Seyfert objects. If sample selection is not the
explanation, then galaxy assembly bias could reconcile these
results---as discussed in the introduction, halos with more
substructure tend to have weaker clustering at fixed $\mhalo$,
implying that late-forming halos are more likely to undergo AGN
activity. Further study is required to reconcile these results.

A third method of obtaining halo massess---distinct from both
clustering and $\lsat$---is galaxy group
catalogs. \cite{weston_etal:17} use a similar WISE color selection to
find dusty AGN in the SDSS MGS for the purpose of investigating merger
activity. They obtain halo masses from the group catalog of
\cite{yang_etal:07_catalog}. When restricting to a sample of
interacting and merging galaxies, the same of WISE AGN is only 18
systems. Within this sample, the median halo mass for central galaxies
is 0.2 dex {\it smaller} than that of the non-AGN sample. However,
given the limited sample size, a KS test yields a $P_{KS}$-value of
0.1. Further, the Yang catalog---along with most other group-finding
algorithms---use the total stellar mass of the group to match to halo
mass. Thus, when the central galaxy dominates the stellar mass of the
group---as it does at the mass scales probed here (e.g.,
\citealt{leauthaud_etal:12_total})---the group finder can yield biases
when assessing the halo masses of groups with similar $\mgal$ values
for the central galaxy (\citealt{campbell_etal:15}). The $\lsat$
method is optimal at differentiating halo masses for these types of
samples.

Do the two AGN samples reside in different halos? In a strict
interpretation of that question, WISE-selected AGN live in halos
roughly twice as massive as BPT-selected AGN. The distribution of
$\mgal$ for both AGN samples peaks at $10^{10.8}$ $\msol$. At this
stellar mass, the mean halo mass is $\sim 10^{12.2}$ $\hmsol$ (cf.,
the compilation of SHMR results in Figure 2 of
\citealt{wechsler_tinker:18}). At this halo mass scale, $\lsat$
corresponds nearly linearly to $\mhalo$.  Thus, under the
interpretation that the differences in $\lsat$ are driven by changes
in $\mhalo$, the $\chibar$ results from Figure \ref{lsatbar} imply
halo masses for the two AGN samples of
$\log M_{\rm WISE}=12.58\pm 0.11$ and
$\log M_{\rm BPT}= 12.31 \pm 0.04$. The difference between the two
halo masses, $0.27 \pm 0.12$ dex, is consistent within 1$\sigma$ with
the results of \cite{dipompeo_etal:17} who find
$\Delta \log \mhalo = 0.45 \pm 0.13$ dex.

However, half of the difference in $\mhalo$ between the two AGN
samples can be explained by differences in the host galaxy population.
In Figure \ref{sample_properties}, the stellar mass and color
distributions of the two AGN populations are similar, but the $\dn$
values are clearly distinct. Most BPT AGN reside in the transition
region between star-forming and quiescent populations. Given the
limited statistics of the WISE sample, once the differences $\dn$ are
controlled, we cannot rule out the null hypothesis that the difference
in $\lsat$ values is driven entirely by the difference in host galaxy
properties.

The host galaxy properties themselves offer additional insight.  If
orientation angle is the only thing that separates dusty AGN from the
rest of the population, then the properties of their host galaxies
should be the same. Now, the $\dn$ distributions in Figure
\ref{sample_properties} by themselves are not indication of a
fundamental difference in the two populations of AGN; rather, it may
indicate selection bias in the construction of the
samples. \cite{mendez_etal:16} also find that IR-selected AGN have a
bias toward star-forming galaxies. \cite{kauffmann_etal:03}
demonstrated that the AGN luminosity, quantified through the
luminosity on the [OIII] emission line, correlates with $\dn$ such
that stronger AGN have younger stellar populations. The left-hand side
of Figure \ref{oiii} shows the distributions of $\log \lumoiii$ in the
BPT sample and the subsample that are also contained within the WISE
selection box in Figure \ref{agn_selection}. Dusty AGN are much
stronger AGN than the typical population. Does this account for the
differences in their $\dn$ values?  The right-hand side of Figure
\ref{oiii} shows that the AGN luminosity accounts for part of the
difference, but cannot fully explain the difference in the host galaxy
distributions. The red and blue histograms reproduce the $\dn$
distributions from Figure \ref{sample_properties}. The solid curve
shows the distribution for the BPT sample, weighted to reproduce the
distribution of $\log \lumoiii$ for the WISE subsample. Although
controlling for luminosity brings the $\dn$ distributions in closer
agreement---just as controlling for $\dn$ brings the $\lsat$ values
closer---they are still clearly distinct distributions in $\dn$.

Future low-redshift data will aide in addressing these questions. The
Bright Galaxy Survey of the Dark Energy Spectroscopic Instrument
(DESI-BGS; \citealt{desi_fdr}) will produce an $r$-band limited
spectroscopic survey complete to $r=19.5$, roughly two magnitudes
fainter than the MGS. This will yield a sample similar in
characteristics to the MGS, but larger by volume and number by a factor
of 20. This will increase the statistics such that the mean
$\lsat$-$\mgal$ relation, as in Figure \ref{lsatbar}, will be a highly
discriminating diagnostic. Measurements of $\lsat$ as a function of
host halo and AGN properties will provide the leverage necessary to
better understand the relationship between halo growth, galaxy growth,
and growth of their supermassive black holes in the local universe.

\section*{Acknowledgements}
The Legacy Surveys consist of three individual and complementary projects: the Dark Energy Camera Legacy Survey (DECaLS; NOAO Proposal ID \# 2014B-0404; PIs: David Schlegel and Arjun Dey), the Beijing-Arizona Sky Survey (BASS; NOAO Proposal ID \# 2015A-0801; PIs: Zhou Xu and Xiaohui Fan), and the Mayall z-band Legacy Survey (MzLS; NOAO Proposal ID \# 2016A-0453; PI: Arjun Dey). DECaLS, BASS and MzLS together include data obtained, respectively, at the Blanco telescope, Cerro Tololo Inter-American Observatory, National Optical Astronomy Observatory (NOAO); the Bok telescope, Steward Observatory, University of Arizona; and the Mayall telescope, Kitt Peak National Observatory, NOAO. The Legacy Surveys project is honored to be permitted to conduct astronomical research on Iolkam Du'ag (Kitt Peak), a mountain with particular significance to the Tohono O'odham Nation.

NOAO is operated by the Association of Universities for Research in Astronomy (AURA) under a cooperative agreement with the National Science Foundation.

This project used data obtained with the Dark Energy Camera (DECam), which was constructed by the Dark Energy Survey (DES) collaboration. Funding for the DES Projects has been provided by the U.S. Department of Energy, the U.S. National Science Foundation, the Ministry of Science and Education of Spain, the Science and Technology Facilities Council of the United Kingdom, the Higher Education Funding Council for England, the National Center for Supercomputing Applications at the University of Illinois at Urbana-Champaign, the Kavli Institute of Cosmological Physics at the University of Chicago, Center for Cosmology and Astro-Particle Physics at the Ohio State University, the Mitchell Institute for Fundamental Physics and Astronomy at Texas A\&M University, Financiadora de Estudos e Projetos, Fundacao Carlos Chagas Filho de Amparo, Financiadora de Estudos e Projetos, Fundacao Carlos Chagas Filho de Amparo a Pesquisa do Estado do Rio de Janeiro, Conselho Nacional de Desenvolvimento Cientifico e Tecnologico and the Ministerio da Ciencia, Tecnologia e Inovacao, the Deutsche Forschungsgemeinschaft and the Collaborating Institutions in the Dark Energy Survey. The Collaborating Institutions are Argonne National Laboratory, the University of California at Santa Cruz, the University of Cambridge, Centro de Investigaciones Energeticas, Medioambientales y Tecnologicas-Madrid, the University of Chicago, University College London, the DES-Brazil Consortium, the University of Edinburgh, the Eidgenossische Technische Hochschule (ETH) Zurich, Fermi National Accelerator Laboratory, the University of Illinois at Urbana-Champaign, the Institut de Ciencies de l'Espai (IEEC/CSIC), the Institut de Fisica d'Altes Energies, Lawrence Berkeley National Laboratory, the Ludwig-Maximilians Universitat Munchen and the associated Excellence Cluster Universe, the University of Michigan, the National Optical Astronomy Observatory, the University of Nottingham, the Ohio State University, the University of Pennsylvania, the University of Portsmouth, SLAC National Accelerator Laboratory, Stanford University, the University of Sussex, and Texas A\&M University.

BASS is a key project of the Telescope Access Program (TAP), which has been funded by the National Astronomical Observatories of China, the Chinese Academy of Sciences (the Strategic Priority Research Program "The Emergence of Cosmological Structures" Grant \# XDB09000000), and the Special Fund for Astronomy from the Ministry of Finance. The BASS is also supported by the External Cooperation Program of Chinese Academy of Sciences (Grant \# 114A11KYSB20160057), and Chinese National Natural Science Foundation (Grant \# 11433005).

The Legacy Survey team makes use of data products from the Near-Earth Object Wide-field Infrared Survey Explorer (NEOWISE), which is a project of the Jet Propulsion Laboratory/California Institute of Technology. NEOWISE is funded by the National Aeronautics and Space Administration.

The Legacy Surveys imaging of the DESI footprint is supported by the Director, Office of Science, Office of High Energy Physics of the U.S. Department of Energy under Contract No. DE-AC02-05CH1123, by the National Energy Research Scientific Computing Center, a DOE Office of Science User Facility under the same contract; and by the U.S. National Science Foundation, Division of Astronomical Sciences under Contract No. AST-0950945 to NOAO.

\bibliography{risa}

\label{lastpage}

\end{document}